\documentclass[pra,twocolumn,amsmath]{revtex4}
\usepackage{graphicx}
\usepackage{amssymb}

\begin{document}

\def\ket#1{|#1\rangle}
\def\bra#1{\langle#1|}
\def\av#1{\langle#1\rangle}
\def\myarrow{\mathop{\longrightarrow}}

\title{Impossibility of large phase shifts via the ``giant Kerr effect'' with single-photon wavepackets}

\author{Julio Gea-Banacloche}
\affiliation{Department of Physics, University of Arkansas, Fayetteville, AR 72701}

\date{\today}

\begin{abstract}
An approximate analytical solution is presented, along with numerical calculations, for a system of two single-photon wavepackets interacting via an ideal, localized Kerr medium.  It is shown that, because of spontaneous emission into the initially unoccupied temporal modes, the cross-phase modulation in the Schr\" odinger picture is very small as long as the spectral width of the single-photon pulses is well within the medium's bandwidth.  In this limit, the Hamiltonian used can be derived from the ``giant Kerr effect'' for a four-level atom, under conditions of electromagnetically-induced transparency; it is shown explicitly that the linear absorption in this system increases as the pulse's spectral width approaches the medium's transparency bandwidth, and hence, as long as the absorption probability remains small, the maximum cross-phase modulation is limited to essentially useless values.  These results are in agreement with the general, causality- and unitarity-based arguments of Shapiro and co-workers.     
\end{abstract}
\maketitle

\section{Introduction}
It was suggested by Chuang and Yamamoto \cite{chuang} that the cross-Kerr effect (or cross-phase modulatation) between two optical fields could be used for so-called ``dual-rail'' quantum logic with photonic qubits, provided sufficiently large nonlinearities could be generated.  The goal is to be able to perform a transformation like the following (where $0$ and $1$ refer to the number of photons in the two interacting modes):
\begin{equation}
\frac{1}{\sqrt 2}\bigl(\ket 0 + \ket 1\bigr)\ket 1 \to \frac{1}{\sqrt 2}\bigl(\ket 0 + e^{-i\phi}\ket 1\bigr)\ket 1
\label{wanted}
\end{equation}
For cross-phase modulation by a $\chi^{(3)}$ medium, the phase shift $\phi = \kappa n_1 n_2$, where $n_1$ and $n_2$ are the number of photons in the respective modes, and $\kappa$ is a constant.  If $\kappa$ can be made as large as $\pi$, then Eq.~(\ref{wanted}) shows that we have a conditional logical operation which, in the basis of states $(\ket 0 \pm \ket 1)/\sqrt 2$, is equivalent to the CNOT gate.

As already pointed out in \cite{nielsen}, however, the very large Kerr nonlinearites needed to achieve this sort of phase modulation are, in ordinary materials, always associated with large absorption losses.  Over the years, a number of methods have been suggested to overcome this difficulty, mostly revolving around the use of electromagnetically-induced transparency, or EIT, to eliminate the linear absorption and, at the same time, increase the nonlinear dispersion of the medium \cite{munro,ottaviani,sanders}.  Nevertheless, in 2006, J. H. Shapiro published a study \cite{shapiro} showing that the causal, non-instantaneous behavior of any $\chi^{(3)}$ nonlinearity would always prevent the transformation (\ref{wanted}) from happening with high fidelity for large $\phi$.  Central to Shapiro's study was a multimode treatment of the two propagating one-photon wavepackets, something that most previous studies had not considered.  

Shapiro's original argument was very general, and in a later publication \cite{shapiro2} it was stated that it was not immediately apparent whether it applied to the EIT schemes.  One such scheme that had, in fact, received a multimode quantum-field treatment was that of Lukin and Imamo\u glu \cite{lukin} (see also the discussion in \cite{rmp}), based on the so-called ``giant Kerr effect,'' originally proposed by Schmidt and Imamo\u glu \cite{ima} (which is, in one form or another, at the heart of all the EIT proposals).  The conclusions of \cite{lukin,rmp} are, however, somewhat ambiguous, because they suggest that very large phase shifts are possible, but at the cost of large changes to the modal structure of the pulses.

The goal of the present paper is to show that with an appropriate, idealized, but local Hamiltonian, that reproduces the Heisenberg-picture evolution equations of refs. \cite{lukin,rmp}, large phase shifts in the Schr\" odinger picture (as in Eq.~(\ref{wanted})) are, in fact, impossible, if the traveling fields are described by localized, single-photon pulses.  This is in agreement with Shapiro's prediction for the ``fast nonlinearity'' case.  In fact, the analysis presented here clearly shows that a large phase shift is only possible in the limit in which the pulse's spectral width approaches the bandwidth of the nonlinear medium.  In that case, however, it is also shown here, for the specific atomic configuration considered \cite{ima} in the derivation of the giant Kerr effect, that EIT becomes ineffective and the losses due to absorption in the medium approach unity.  

The paper is organized as follows.  The instantaneous response limit, with an idealized Hamiltonian, is considered in Section 2.  The actual time response of an EIT medium is considered in Section 3, where a relationship between absorption losses and the pulse's spectral width is derived.  Section 4 contains some further discussion and conclusions.

\section{Instantaneous response limit}
\subsection{Hamiltonian, locality, and Heisenberg picture results}
Suppose that, somehow, one has managed to produce a medium that leads, to a sufficiently good approximation, to a field evolution described by the following Hamiltonian
\begin{align}
H = &\sum_{n=-n_{max}}^{n_{max}} \hbar \frac{2\pi n c}{L} a^\dagger_n a_n + \sum_{m=-n_{max}}^{n_{max}} \hbar \frac{2\pi m c}{L} b^\dagger_m b_m \cr
&+ \hbar\epsilon \int_{z_0}^{z_0+l}E_a^{(-)} E_a^{(+)} E_b^{(-)} E_b^{(+)} \, dz
\label{ham}
\end{align}
Here $E_a$ and $E_b$ are Schr\" odinger-picture field operators given by 
\begin{align}
E_a^{(+)} &= \left(\frac{\hbar\omega_0}{\epsilon_0 A L}\right)^{1/2}\sum_{n=-n_{max}}^{n_{max}} a_n e^{2 \pi i n z/L} \cr
E_b^{(+)} &= \left(\frac{\hbar\omega_0}{\epsilon_0 A L}\right)^{1/2}\sum_{m=-n_{max}}^{n_{max}} b_m e^{2 \pi i m z/L}
\end{align}
with $E_a^{(-)}, E_b^{(-)}$ their Hermitian conjugates.  The quantization volume has cross-sectional area $A$ and length $L$.  The total number of modes to be added, for each field, is $M=2 n_{max} + 1$, and it is determined by $L$ and the bandwidth of the optical medium, ${\Delta\omega}_\text{medium} = 2 \pi c M/L$.  The number of initially-occupied field modes may be much less than $M$, as will be discussed below.  The following Section discusses how the ``giant Kerr effect'' \cite{ima} may lead to the kind of interaction Hamiltonian that appears in (\ref{ham}) under appropriate conditions.

An important property of the Hamiltonian (\ref{ham}) is that it is \emph{local}. Suppose that the field state is described by a pair of wavepackets, which for simplicity we will take to be identical:
\begin{equation}
\ket{\psi_0} = \sum_{n=-n_{max}}^{n_{max}} c_n \ket{1_n}_a \otimes \sum_{n=-n_{max}}^{n_{max}} c_n \ket{1_n}_b
\end{equation}
Here $\ket{1_n}_a$ is a state with one photon in the $n$-th ``a'' mode, and zero photons in all the other modes, and similarly $\ket{1_n}_b$.  The field intensity in such a state is given by
\begin{equation}
I_a(z) = \av{E_a^{(-)} E_a^{(+)}} = \frac{\hbar\omega_0}{\epsilon_0 A L} \left|\sum_n c_n e^{2 \pi i n z/L}\right|^2
\label{iaz}
\end{equation}
and the expectation value of the energy is
\begin{equation}
\av{H} = \frac{4\pi\hbar c}{L} \sum n |c_n|^2 + \hbar\epsilon \int_{z_0}^{z_0+l} I_a(z) I_b(z) \, dz
\label{avh}
\end{equation}
For pulses with a symmetric spectrum, such as those which will be considered here, the first term in (\ref{avh}) vanishes. The second term, on the other hand, which represents the interaction energy between the field and the material medium, vanishes if (and only if) the pulses are not inside the medium.  This is a physically reasonable requirement for any Hamiltonian that one might want to use to describe the interaction of a finite pulse with a localized medium.

If the Heisenberg equation of motion is used with the Hamiltonian (\ref{ham}), one gets the following propagation equations for the field operators $E_a^{(+)}(t,z)$ and $E_b^{(-)}(t,z)$, in the Heisenberg picture:
\begin{align}
\left(\frac{\partial}{\partial t} + c \frac{\partial}{\partial z}\right)E_a^{(+)} &=  i \kappa E_b^{(-)} E_b^{(+)} E_a^{(+)} \\
\left(\frac{\partial}{\partial t} + c \frac{\partial}{\partial z}\right)E_b^{(+)} &=  i \kappa E_a^{(-)} E_a^{(+)} E_b^{(+)}
\label{heis}
\end{align}
where $\kappa = \epsilon(\hbar\omega_0/\epsilon_0 A)$, and the right-hand side is as shown for $z_0\le z\le z_0 + l$, and $0$ elsewhere.  This follows from the assumption that the bandwidth of the medium is large enough to justify the approximation
\begin{equation}
\sum_{n=-n_{max}}^{n_{max}} e^{2\pi in(z-z^\prime)/L} \simeq = L\delta(z-z^\prime)
\end{equation}
Strictly speaking, this requires also that the bandwidth of the pulses be smaller than the bandwidth of the interaction, an assumption to which we will return shortly.

Eqs.~(\ref{heis}) are exactly the same as obtained by Lukin and Imamo\u glu in \cite{lukin}, minus all the extra complications arising from the different group velocities in the medium and the wavepacket compression.  These complications will simply be ignored here in order to concentrate on the basic difficulties caused by locality and the multimode nature of the field.  As explained in \cite{lukin}, the Eqs.~(\ref{heis}) can be self-consistently solved by integrating along the characteristics to get, at any point $z$ within the medium
\begin{align}
E^{(+)}_{a,b}(t,z) = &E^{(+)}_{a,b}(t^\prime,z_0) \cr
&\times\exp\left[i\frac{\kappa}{c} (z-z_0) E^{(-)}_{b,a}(t^\prime,z_0) E^{(+)}_{b,a}(t^\prime,z_0) \right]\cr
\label{heissol}
\end{align}
with the local time $t^\prime \equiv t -(z-z_0)/c$.  Eq.~(\ref{heissol}) can be verified by direct substitution in (\ref{heis}), noting that it implies the equality $E^{(-)}_{a,b}(t,z)E^{(+)}_{a,b}(t,z) = E^{(-)}_{a,b}(t^\prime,z_0)E^{(+)}_{a,b}(t^\prime,z_0)$.  For $z>z_0+l$, one can use free propagation backwards, and the fact that the field only undergoes multiplication by a unitary operator, to rewrite the result in terms of the $t=0$ operators:
\begin{align}
E^{(+)}_{a,b}(t,z) = &E^{(+)}_{a,b}(0,z-ct) \cr
&\times\exp\left[i\frac{\kappa l}{c} E^{(-)}_{b,a}(0,z-ct) E^{(+)}_{b,a}(0,z-ct) \right]\cr
\label{heissol2}
\end{align}

At first sight, Eq.~(\ref{heissol2}) might seem to be exactly what we want, since it suggests that each of the two fields acquires a phase that is proportional to the intensity of the other one.  The fact that the actual phase shift apparently depends on the local intensity, at different points in the wavepacket, may be slightly worrisome, but Eq.~(\ref{heissol2}) at least suggests that nothing should prevent one from making the phase at, say, the center of the wavepacket, as large as one might want to.  The situation looks quite different, however, in the Schr\" odinger picture, to which we turn next.

\subsection{Time evolution in the Schr\" odinger picture}

In the Schr\" odinger picture we write the state of the system as the double sum
\begin{equation}
\ket{\psi(t)} = \sum_n \sum_m c_{nm}(t) e^{-2 \pi i(n+m)c t/L} \ket{1_n}_a\ket{1_m}_b
\end{equation}
where the coefficient $c_{nm}(0)$ (two indices) equals the product $c_n(0)c_m(0)$ (single index) at $t=0$.  The equation of motion for $c_{nm}$ is
\begin{align}
\dot c_{nm} = &-i\epsilon \left(\frac{\hbar\omega_0}{\epsilon_0 A L} \right)^2\sum_{n^\prime m^\prime}c_{n^\prime m^\prime} \cr
&\times\int_{z_0}^{z_0+l} e^{-2\pi i(n^\prime + m^\prime - n - m)(ct-z)/L}\, dz
\label{dotcnm}
\end{align}
This can be integrated analytically under some approximations that are equivalent to the ones used in the previous section. To begin with, introduce a new set of indices, $\mu$ and $\nu$, that stand for the sum and difference, respectively, of $n$ and $m$.  Then $c_{\nu\mu} = c_{nm}$ with $n=(\mu+\nu)/2$ and $m=(\mu-\nu)/2$, and we have
\begin{equation}
\dot c_{\nu\mu} = -i\eta  \sum_{\mu^\prime}\left(\sum_{\nu^\prime }c_{\nu^\prime \mu^\prime} \right)\int_{z_0}^{z_0+l} e^{-2\pi i(\mu^\prime - \mu)(ct-z)/L}\, dz
\label{dotcnumu}
\end{equation}
where, for convenience, the parameter $\eta = \epsilon(\hbar\omega_0/\epsilon_0 A L)^2$ has been defined.  One can next introduce a new set of coefficients, $v_\mu(t)$, defined by
\begin{equation}
v_\mu = \sum_{\nu=|\mu|-2 n_{max}}^{2 n_{max}-|\mu|} c_{\nu\mu}
\label{defv}
\end{equation}
Note that in (\ref{defv}), as in all other sums over $\nu$ for constant $\mu$, the index $\nu$ increases in steps of 2, so there are $2n_{max} -|\mu|+1$ terms in the sum (and $\mu$ ranges from $-2n_{max}$ to $2 n_{max}$).  The $v_\mu$ obey the equation of motion (easily derived from (\ref{dotcnumu}))
\begin{align}
\dot v_\mu = &-i\eta \left(2n_{max}- |\mu|+1 \right) \sum_{\mu^\prime=-2 n_{max}}^{2 n_{max}}  v_{\mu^\prime} \cr
&\times\int_{z_0}^{z_0+l} e^{-2\pi i(\mu^\prime - \mu)(ct-z)/L}\, dz
\label{dotvmu}
\end{align}
Equation (\ref{dotvmu}) can be integrated by introducing an envelope function $f(t,z)$, defined as
\begin{equation}
f(t,z) = \sum_\mu v_\mu(t) e^{-2\pi i\mu \omega(ct-z)/L}
\label{deff}
\end{equation}
which satisfies
\begin{widetext}
\begin{align}
\left(\frac{\partial}{\partial t} + c \frac{\partial}{\partial z}\right)f &= \sum_\mu \dot v_\mu e^{-2\pi i\mu \omega(ct-z)/L} \cr
&= - i\eta \sum_{\mu^\prime} v_{\mu^\prime} \int_{z_0}^{z_0+l} e^{-2\pi i\mu^\prime \omega(ct-z^\prime)/L} \left[\sum_\mu (2n_{max}-|\mu|+1)e^{2 \pi i\mu (z-z^\prime)/L}\right] dz^\prime
\label{partialsf}
\end{align}
\end{widetext}
It is easy to see that, for large enough $n_{max}$, the expression in square brackets in (\ref{partialsf}) converges to $(2 n_{max} + 1)L\delta(z-z^\prime)$, which means that we have
\begin{equation}
\left(\frac{\partial}{\partial t} + c \frac{\partial}{\partial z}\right)f = -i\eta M L f(t,z), \qquad z_0 < z < z_0 + l
\label{partialsf2}
\end{equation}
where, again, $M=2n_{max} + 1$ is the total number of modes, and the right-hand side of (\ref{partialsf2}) vanishes outside the medium.  Integrating along the characteristics, as in the previous section, one finds, for $z$ inside the medium,
\begin{align}
f(t,z) &= e^{-i\eta ML (z-z_0)/c}f(0,z-ct) \cr
&= e^{-i\eta M L (z-z_0)/c} \sum_\mu v_\mu(0) e^{-2\pi i\mu (c t-z)/L}
\end{align}
which can now be substituted into (\ref{dotcnumu}) (via the definitions (\ref{defv}) and (\ref{deff})), to yield
\begin{align}
\dot c_{\nu\mu} = &-i\eta \sum_{\mu^\prime} v_{\mu^\prime}(0) \cr
&\times\int_{z_0}^{z_0+l} e^{-2 \pi i(\mu^\prime-\mu) (ct-z)/L}  e^{-i\eta ML (z-z_0)/c}  \, dz \cr
\label{dotcnumu2}
\end{align}
The expression on the right-hand side of (\ref{dotcnumu2}) can now be directly integrated. For the purpose of comparing the state of the field after the interaction to the state before the interaction, it is convenient to concentrate on the value of the coefficients $c_{\nu\mu}$ at the time $T\equiv L/c$ (i.e., the quantization time), at which point, in the absence of interaction, the pulse should return to its initial state, since the traveling-wave formalism we are using is equivalent to periodic boundary conditions.  In that case, the integration of (\ref{dotcnumu2}) over time from $t=0$ to $t=T$ selects only the $\mu^\prime = \mu$ term in the sum, and we have
\begin{equation}
c_{\mu\nu}(T) = c_{\mu\nu}(0) + \frac{1}{M}\left(e^{-i\eta MLl/c}-1 \right)\, v_{\mu}(0)
\label{cmunut}
\end{equation}

\subsection{Fidelities, and numerical results}  
Equation (\ref{cmunut}) can be used to calculate the overlap between the initial state and the state at the time $T$, from which, in turn, a number of other useful results can be derived.  Defining, for simplicity, 
\begin{equation}
\Phi = \frac{\eta MLl}{c} = \frac{\kappa l}{c} M \frac{\hbar\omega_0}{\epsilon_0 AL}
\end{equation}
(where the last expression uses $\kappa$ as defined in the previous section, for comparison with the Heisenberg-picture result, Eq.~(\ref{heissol2})), we have
\begin{equation}
\av{\psi(0)|\psi(T)} = 1 + \frac{1}{M} \left(e^{-i\Phi}-1\right) \sum_\mu |v_\mu(0)|^2 \equiv \sqrt{{\cal F}_0}\,e^{-i\phi}
\label{proj}
\end{equation}
Recall that the original goal (Eq.~(\ref{wanted})) was to leave the original two-photon state invariant except for a phase shift.  The fidelity ${\cal F}_0$ is a measure of the success of this operation.  Note that if $\sum_\mu|v_\mu(0)|^2/M =1$, we have ${\cal F}_0 = 1$ and $\phi=\Phi$.  It is important, therefore, to calculate this quantity.  Note that the expression (\ref{iaz}) for the single-wavepacket intensity can be rewritten as
\begin{equation}
I_a(z) = \av{E_a^{(-)}E_a^{(+)}} = \frac{\hbar\omega_0}{\epsilon_0 A L} \sum_{n,m} c_n^\ast(0) c_m(0) e^{-2\pi i(n-m)z/L} 
\end{equation}
from which it follows that
\begin{equation}
I_a^2(z) = \left(\frac{\hbar\omega_0}{\epsilon_0 A L}\right)^2 \sum_{\mu,\mu^\prime} v_\mu^\ast v_{\mu^\prime} e^{-2 \pi i(\mu-\mu^\prime)z/L}
\end{equation}
and therefore
\begin{equation}
\sum_\mu |v_\mu(0)|^2 = \frac 1 L \left(\frac{\epsilon_0 A L}{\hbar\omega_0}\right)^2\,\int_0^L I_a^2(z)\, dz = \frac{L \int_0^L I_a^2(z)\, dz}{\left[\int_0^L I_a(z)\, dz \right]^2}
\label{dasum}
\end{equation}
This can be related to the pulse bandwidth as follows.  First, note that the assumption of a localized pulse means that it is legitimate to extend all the integral arguments in (\ref{dasum}) from minus infinity to infinity.  Then, introduce the spatial Fourier transform $P(k)$ of the function $I_a(z)$, so that
\begin{equation}
I_a(z) = \frac{1}{\sqrt{2\pi}} \int_{-\infty}^\infty P(k) e^{ikz} dk
\end{equation}
Then we have
\begin{equation}
\sum_\mu |v_\mu(0)|^2 = \frac{L}{2\pi} \frac{\int_{-\infty}^{\infty} |P(k)|^2 dk}{|P(0)|^2}
\end{equation}
Now suppose that the function $P(k)$ is peaked at $k=0$ (as it should be if we have correctly separated the slowly-varying part of the pulse form its carrier frequency), and that it is negligible outside of an interval of width $\Delta k$.  Then $|P(k)|^2/|P(0)|^2 \le 1$ for all $k$, and therefore
\begin{equation}
\sum_\mu |v_\mu(0)|^2 \le \frac{L}{2\pi} \Delta k
\label{dasum2}
\end{equation}
where the equality holds only for a ``square'' $P(k)$ (constant in the interval $\Delta k$, and zero outside of it); but this corresponds to a spatially non-localized pulse, whose intensity decays only as $1/z$.  We can then assume that (\ref{dasum2}) is always a strict inequality.  Returning to our original formulation in terms of $M$ discrete modes spaced, in frequency, by $2\pi c/L$, we conclude that
\begin{equation}
r\equiv \frac 1 M \sum_\mu |v_\mu(0)|^2 < \frac{\Delta\omega_\text{pulse}}{\Delta\omega_\text{medium}}
\label{rdef}
\end{equation}
which should always be less than 1 (note that $\Delta\omega_\text{pulse}$ has been defined through the effective support of $P(k)$, and will typically be larger than the conventional ``standard deviation'' of $\omega$ for the wavepackets considered).  In terms of the quantity $r$, we can express the fidelity ${\cal F}_0$ and the phase $\phi$ as
\begin{equation}
{\cal F}_0 = 1 - 4 \sin^2\left(\frac \Phi 2\right) \, r(1-r)
\label{deff0}
\end{equation}
and
\begin{equation}
\phi = \tan^{-1}\left[\frac{r \sin\Phi}{1-r + r\cos\Phi}\right]
\label{defphi}
\end{equation}
Note that we can always make the distortion of the original wavepacket negligible by letting $r\to 0$, but this is at the expense of an extremely small phase shift.  The opposite case, $r\to 1$, also leads to a high fidelity, this time with potentially large phase shifts, but it is forbidden by locality.  By this it must be understood that it is simply not possible to make $r$ arbitrarily close to 1 with a localized pulse.  For instance, suppose that the intensity $I_a$ is proportional to a Gaussian $e^{-(z-z_1)^2/\sigma^2}$.  The right-hand side of (\ref{dasum}) then evaluates to $1/(\sigma\sqrt{2\pi})$, and we have
\begin{equation}
r = \frac{L/\sigma}{M} \frac{1}{\sqrt{2\pi}} < 0.4 \qquad \text{(Gaussian)}
\label{rgaussian}
\end{equation}
since, in order to describe the pulse adequately, the number of modes $M$ must be at least of the order of $L/\sigma$ (in other words, using fewer modes results in a non-localized pulse).  Similarly, for a hyperbolic secant $I_a(z) \sim \text{sech}(z/\sigma)$, we find
\begin{equation}
r = \frac{L/\sigma}{M} \frac{2}{\pi^2} < 0.2 \qquad \text{(hyperbolic secant)}
\end{equation}
Since order of magnitude arguments are always uncertain regarding such things as factors of two, however, and the temptation to look for an ``optimal pulse shape'' is strong, it is important to keep in mind the absolute bound (\ref{rdef}), and also the considerations to follow in the next section, which will show, for the specific example of a ``giant Kerr effect'' medium, how things deteriorate when one tries to fit the pulse's spectrum too tightly in the medium's transparency window. 

Parametric plots of ${\cal F}_0$ versus $\phi$ for various values of $r<0.5$ are shown in Figure 1.  Note that, when $r=1/2$, $\phi = \Phi/2$ and ${\cal F}_0 = \cos^2(\Phi/2)$. 
\begin{figure}
\begin{center}
\includegraphics[width=3.2in]{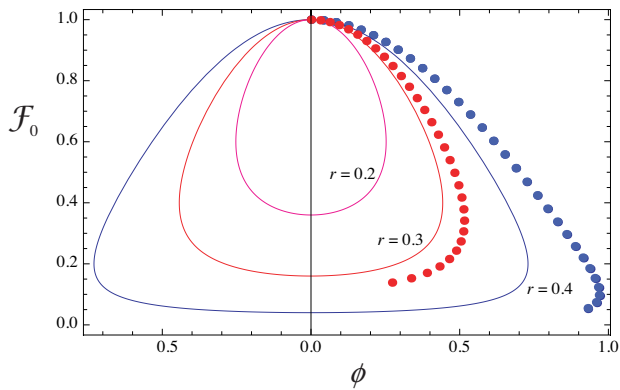}
\end{center}
\caption[example]
   { \label{fig:fig1}
Parametric plot of the fidelity ${\cal F}_0$ versus the phase shift $\phi$ for $r=0.4, 0.3, 0.2$, as $\Phi$ is varied from $0$ to $2\pi$.  Dots: result of numerical calculations with a Gaussian pulse, with 17 modes and $\sigma/L = 0.059$ (corresponding to $r=0.4$, outermost dots), and $\sigma/L = 0.078$ (corresponding to $r=0.3$, innermost dots); the numerical calculations cover only the range $0\le\Phi\le \pi$.}
\end{figure}
The figure also shows the results of numerical calculations (based on the direct integration of Eq.~(\ref{dotcnm})) for $M=17$ modes and two Gaussian pulses with $I_a \sim e^{-(z-z_1)^2/\sigma^2}$, $z_1 = L/4$, and $\sigma = 0.059 L$ (corresponding to $r=0.4$) and $\sigma=0.078 L$ (corresponding to $r=0.3$).  The nonlinear medium was taken to start at $z_0=L/2$ and had extension $l=L/2$.  The agreement with the analytical approximation is better for the broader pulse because the replacement of the term in square brackets in (\ref{partialsf}) by a delta function is a better approximation in that case.  The calculations show that the theory may underestimate somewhat the achievable phase $\phi$, but large values of $\phi$ still correspond to very small fidelities.

\subsection{Discussion}

The results in the previous subsection indicate that the parameter $r$ that determines the maximum achievable phase shift decreases, as the ratio of the pulse's spectral width to the medium's bandwidth.  

One may wonder why, once enough modes have been included in the calculation to describe the wavepacket properly, the addition of more, empty, modes should have a degrading effect on the performance of the system.  The answer lies in spontaneous emission.  Considering the action of the Hamiltonian (\ref{ham}) on an initial state with only one photon in each pulse, the two sets of annihilation operators, $E^{(+)}_a$ and $E^{(+)}_b$, will first produce a uniform vacuum state, and then the creation operators $E^{(-)}_a$ and $E^{(-)}_b$ may replace the $a$ and $b$ photons into any of the available temporal modes, regardless of where they may have been originally.  There are some momentum and energy conservation constraints, enforced by integrals over position and time respectively, but as (for instance) Eq.~(\ref{dotcnumu}) shows, an initially unoccupied pair of modes with an arbitrary $\nu$ (difference between ``a'' and ``b'' temporal frequency) can be created out of any of the preexisting pairs of modes with the same $\mu$ (sum of the ``a'' and ``b'' frequencies), without incurring any energy or momentum penalty. This is also apparent from Eq.~(\ref{cmunut}).

This problem is particularly acute for single-photon wavepackets.  For pulses containing an appreciable number of photons, $\bar n$, the action of the annihilation operators results in modes that are still highly populated, and so stimulated emission will take place preferentially in those modes.  In other words, one expects this problem to decrease as $1/\bar n$ (the ratio of spontaneous to stimulated emission), as the number of photons in the pulses increases.

To get the single-photon case to work, one might contemplate modifying the Hamiltonian, so that, for instance, instead of the negative-frequency field operators $E^{(-)}_a$ and $E^{(-)}_b$ one would have a weighted sum of creation operators, more closely matching the spectrum of the incoming pulse (of course, by Hermiticity, the positive-frequency field operators $E^{(+)}_a$ and $E^{(+)}_b$ would also have to be modified).  This amounts to introducing some of the effects of dispersion in the medium, but it cannot be done arbitrarily, since there are physical rules (such as the Kramers-Kronig relations) that govern these things.  In particular, strong dispersion is typically associated with absorption.  As will be shown in the next section, even the extremely weak residual absorption present in the giant (EIT-enhanced) Kerr effect is enough to prevent one from taking the limit $r\to 1$ in the results presented above.

It may be worth considering for a moment the extreme case of a ``toy'' Hamiltonian that would work with any pulse shape.  This could be achieved by replacing the interaction part of (\ref{ham}) by
\begin{equation}
H_I = \hbar l \epsilon \left(\sum_n a^\dagger_n a_n\right)\left(\sum_m b^\dagger_m b_m\right)
\label{hnonlocal}
\end{equation}
Unlike (\ref{ham}), this Hamiltonian does not create any photons in initially unoccupied modes; yet, it is unphysical, because it is completely nonlocal: the pulse will be interacting with the medium wherever it might happen to be.  This highlights the importance of doing multimode quantized-field calculations properly.  Formally, either one of the Hamiltonians (\ref{ham}) or (\ref{hnonlocal}) could be considered as a possible generalization of the single-mode Kerr Hamiltonian $a^\dagger a b^\dagger b$, but they yield very different predictions in the multimode case, and only one of them is (approximately) physical.  It is to the terms neglected in this approximation that we turn in the next Section.

\section{Giant Kerr effect and medium bandwidth}

In the previous section it was shown that in order to achieve a relatively large phase shift one should try to make the pulse's bandwidth as close to that of the medium as possible.  However, when one does that, the medium's absorption is not negligible anymore.

\begin{figure}
\begin{center}
\includegraphics[width=3in]{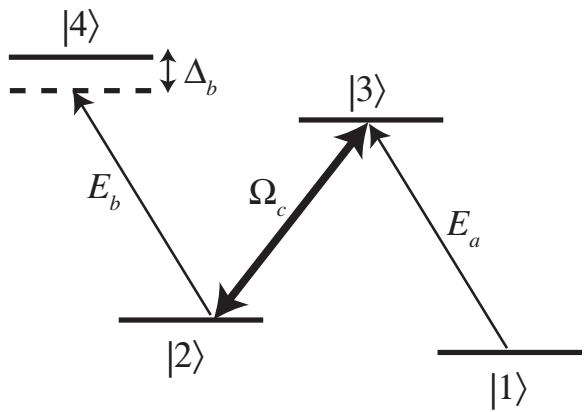}
\end{center}
\caption[example]
   { \label{fig:fig2}
Level scheme for the giant Kerr effect.  $E_a$ and $E_b$ are weak (in this paper, single-photon) fields; $\Omega_c$ is the EIT ``coupling field.''}
\end{figure}

A way to approximately realize the Hamiltonian (\ref{ham}) is by making use of the ``giant Kerr effect'' introduced in \cite{ima}.  The scheme (illustrated in Figure 2) makes use of electromagnetically-induced transparency, or EIT, to open a ``transparency window'' in the medium for the field $E_a$ (the field $E_b$ does not need it, since the level 2 will typically be unoccupied, and the detuning $\Delta_b$ will be taken to be large), as well as to enhance the dispersion of the medium and with it the Kerr nonlinearity.  As shown in \cite{ima}, if the optical Bloch equations for the atom are solved in steady-state, under the assumption that absorption is small, one obtains for the atomic dipole amplitudes in the two transitions the result
\begin{align}
p_a &= i\frac{4 d_{13}^2 d_{24}^2 }{\hbar^4\Omega_c^2(\gamma_4/2 +i \Delta_b)}\,|E_b|^2 E_a^\ast \label{e18b} 
\\
p_b &= i\frac{4 d_{13}^2 d_{24}^2 }{\hbar^4\Omega_c^2(\gamma_4/2 +i \Delta_b)}\,|E_a|^2 E_b^\ast \label{e18a} \end{align}
where $d_{13}$ and $d_{24}$ are the dipole matrix elements for the two transitions.  Multiplying each dipole amplitude by the corresponding field, and adding the contributions of all the atoms by integrating over the spatial extent of the medium, then yields an interaction energy of the form (\ref{ham}), with 
\begin{equation}
\epsilon = \frac{4 d_{13}^2 d_{24}^2}{\hbar^4\Delta_b\Omega_c^2} \rho A
\end{equation}
where $\rho$ is the volume density of atoms in the medium and $A$ the cross-sectional area of the beam, and the assumption $\Delta_b \gg \gamma_4$ has been made.  As discussed in, e.g., \cite{rmp}, this steady-state approximation, appropriate for a continuous-wave field, neglects a number of important dispersive effects that result in a slowing down and broadening of the $E_a$ pulse. These complications were discussed by Harris and Hau in \cite{harrishau}, and possible ways around them were suggested by Lukin and Imamo\u glu in \cite{lukin}.  Here the propagation effects will be ignored, in order to concentrate, instead, on the consequences of the \emph{temporal} variation of the pulse at the location of each atom.

For an optically dense medium, the transparency window is a Gaussian of width
\begin{equation}
\Delta\omega_\text{trans} = \frac{\Omega_c^2}{\sqrt{\Gamma_{31}\gamma_{31}}}\, \frac{1}{\sqrt{\rho\sigma_a l}}
\label{deltaomegatrans}
\end{equation}
(see, e.g., \cite{lukinscully}) where $\sigma_a = 3\lambda^2/2\pi$ is the on-resonance absorption cross-section of the atom in the $1\to 3$ transition; $\Gamma_{31}$ is the spontaneous emission decay rate from level 3 to level 1, and $\gamma_{31} \ge \Gamma_{31}$ is the total decay rate of the $(1,3)$ coherences, including dephasing and decay to other levels (such as 2).  The residual absorption inside this window has been discussed using a purely semiclassical treatment in \cite{myoptcomm}; here, for single-photon pulses, it will be estimated (and with it, implicitly, the width of the window itself) as follows.  Considering only a one-photon wavepacket in field $a$, and ignoring field $b$ altogether for simplicity, an initial state $\sum_n c_n(0)\ket{1_n}_a\ket 1$ (where the second ket refers to the atomic state) can evolve into a superposition
\begin{equation}
\ket{\psi(t)} = \sum_n c_n(t) e^{-in\omega t}\ket{1_n}_a\ket 1 + C_2(t)\ket{0}_a\ket 2 + C_3(t)\ket{0}_a\ket 3
\end{equation}
(here and in what follows, $\omega \equiv 2\pi c/L$), under the non-Hermitian Hamiltonian
\begin{align}
H_{appr} = &\sum_n\hbar n\omega a^\dagger a - d_{13} \left(E^{(+)}_a \ket 3\bra 1 + E^{(-)}_a \ket 1\bra 3 \right) \cr
&+ \frac{\hbar\Omega_c}{2} \bigl(\ket 2\bra 3 + \ket 3\bra 2\bigr) -i\hbar \frac{\gamma_{31}}{2} \ket 3\bra 3 
\end{align}
For simplicity (in order to use the quasi-pure state approach) we shall neglect the dephasing contribution to $\gamma_{31}$ and the feedback, through spontaneous emission, from state $\ket 3$ to state $\ket 1$.  We still allow for $\Gamma_{31} \ne \gamma_{31}$, and note the relation between the atomic dipole moment matrix element $d_{13}$ and $\Gamma_{31}$:
\begin{equation}
\Gamma_{31} = \frac{\omega_0^3 d_{13}^2}{3 \pi \epsilon_0 \hbar c^3}
\label{G31}
\end{equation}
The equations of motion for the coefficients $c_n, C_3$ and $C_2$ are
\begin{subequations}
\begin{align}
\dot c_n &= i g_{13} C_3 e^{in\omega t} \label{dotcn}\\
\dot C_3 &= -\frac{\gamma_{31}}{2}C_3 + i g_{13}\sum_n c_n e^{-in\omega t} -i \frac{\Omega_c}{2} C_2 \label{dotc3}\\
\dot C_2 &= -i \frac{\Omega_c}{2} C_3 \label{dotc2}
\end{align}
\label{ceqs}
\end{subequations}
where $g_{13}=(d_{13}/\hbar)(\hbar\omega_0/\epsilon_0 AL)^{1/2}$.  Assuming that all the coefficients vary sufficiently slowly, the equation (\ref{dotc3}) for $C_3$ can be adiabatically integrated with the result
\begin{equation}
C_3(t) \simeq i g_{13} \sum_n \frac{c_n(t)e^{-in\omega t}}{\gamma_{31}/2-in\omega} -i\frac{\Omega_c}{\gamma_{31}} C_2(t)
\label{c3ta}
\end{equation}
When this is substituted into the equation (\ref{dotc2}) for $C_2$, the decay rate $\Omega_c^2/2\gamma_{31}$ appears multiplying $C_2$ itself.  Since this rate is, presumably, much greater than the transparency bandwidth (\ref{deltaomegatrans}), it is consistent to assume that all the relevant modes are slower than it, and to perform a further adiabatic integration, with the result
\begin{equation}
C_2(t)\simeq \frac{\Omega_c g_{13}}{2}\sum_n\frac{c_n(t)e^{-in\omega t}}{(\gamma_{31}/2-in\omega)(\Omega_c^2/2\gamma_{31}-in\omega)}
\end{equation}
Finally, this can be substituted again in (\ref{c3ta}), and the lowest-order nonvanishing contribution in $n\omega$ kept, to yield the occupation probability amplitude for level $3$ in the presence of the pulse:
\begin{equation}
|C_3|^2 \simeq \frac{16 g_{13}^2}{\Omega_c^2} \left|\sum_n n\omega c_n e^{-in\omega t}\right|^2
\end{equation}
(note that this is consistent with the semiclassical treatment of \cite{myoptcomm}, which yielded an occupation probability of level 3 proportional to the square of the time-derivative of the field amplitude envelope).

We assume that irreversible processes, represented by the rate $\gamma_{31}$, take the system out of the state 3 and destroy the coherence, and we can estimate then a single-atom ``loss'' probability by
\begin{align}
\int_0^{T} \gamma_{31} |C_{31}|^2 dt &\simeq \frac{16\gamma_{31} g_{13}^2}{\Omega_c^2} \,\frac L c \sum_n (n\omega)^2|c_n(0)|^2 \cr
&= \frac{8\gamma_{31}\Gamma_{31}}{\Omega_c^2}\,\frac{\sigma_a}{A}\,(\delta\omega_\text{pulse})^2
\end{align}
where $T=L/c$ is the ``quantization time,'' Eq.~(\ref{G31}) has been used, and $\delta\omega_\text{pulse}$ (the standard deviation of $n\omega$ for the pulse) has been defined in a natural way.  Multiplying this by the total number of atoms, $\rho A l$, with which the pulse interacts, we obtain the total loss probability,
\begin{equation}
P_\text{loss} = 8 \left(\frac{\delta\omega_\text{pulse}}{\Delta\omega_\text{trans}}\right)^2 \gtrsim  r^2,
\label{ploss}
\end{equation}
assuming that the relevant ``medium bandwidth'' to be used in the calculations in the previous section (in particular, in Eq.~(\ref{rdef})) is of the order of $\Delta\omega_\text{trans}$, and also that the ``effective frequency support'' of the pulse, $\Delta\omega_\text{pulse}$, is of the order of a few standard deviations.  

Eq.~(\ref{ploss}) shows that, in order to prevent the loss of coherence through spontaneous emission out of the level 3, the parameter $r$ needs to be kept very small, in which case, as shown in the previous section, the phase shift in the Schr\" odinger picture is necessarily very small as well.  It may be tempting to try to look for an ``optimal'' pulse shape that, for instance, maximizes ${\cal F}_0$ and $\phi$ (Eqs.~(\ref{deff0}), (\ref{defphi})) while minimizing Eq.~(\ref{ploss}), but that would be missing the point.  The basic meaning of Eq.~(\ref{ploss}) is actually that the unitary evolution under the Hamiltonian (\ref{ham}), assumed in the previous Section, simply does not hold unless the pulse's frequency spectrum is well within the medium's (EIT) transparency bandwidth, in which case $r$, and the maximum phase shift $\phi$, are necessarily small.

As an example, suppose one has a Gaussian pulse of the form $I_a \sim e^{-(z-z_1)^2/\sigma^2}$, in which case $(\delta\omega_\text{pulse})^2 = c^2/2\sigma^2$, and $r$ is given by Eq.~(\ref{rgaussian}).  Then Eq.~(\ref{ploss}) becomes $P_\text{loss} = 2 r^2/\pi$, and to have $P_\text{loss}$ smaller than, say, $0.1$, we require $r \le 0.4$.  If we also want $1-{\cal F}_0 \simeq 0.1$, Eq.~(\ref{deff0}) shows that $\Phi$ cannot exceed $0.66$, and then, by Eq.~(\ref{defphi}), we have $\phi \le 0.26$.  However, this is such a small phase shift that the overlap of the initial state in Eq.~(\ref{wanted}) with the target state is already $\cos^2(\phi/2) = 0.98$.  This means that one has a bigger ``success probability'' if one simply \emph{does nothing at all} to the initial state.

\section{Conclusions}

The results presented here are fully in agreement with the analysis of Shapiro and co-workers \cite{shapiro,shapiro2}.  In particular, in the ``fast nonlinearity'' regime, the achievable phase shift is very small for as long as the instantaneous response approximation is justified.  This corresponds to being allowed to neglect the higher-order (in $n\omega$) terms in the adiabatic expansion in Section 3, which are responsible for the breakdown of unitary evolution as the pulse's bandwidth approaches the EIT transparency bandwidth.  Note that in the formalism used in Section 3 this loss of unitarity is ultimately due to the disappearance of a photon from the system (the photon is absorbed, to bring the atom to level $\ket 3$, and then spontaneously emitted into some other mode); if this was to be described using field operators, restricted to only the two sets of modes ``a'' and ``b'', one would have to throw in a Langevin noise term to preserve the commutation relations.  This would connect to Shapiro's explanation of the reduced fidelity in this regime in terms of phase noise.  (The connection is, essentially, the fluctuation-dissipation theorem.)

A somewhat surprising result from the analysis in Section 2 is the realization that the seemingly arbitrarily large phase obtained in the Heisenberg picture does not necessarily translate into a large phase in the Schr\" odinger picture.  This highlights an important feature of the multimode calculations.  For a single mode it is certainly the case that any phase factor acquired by the Heisenberg-picture operators $a(t)$, $b(t)$, will also appear multiplying the single-photon state $\ket{11}$ in the Schr\" odinger picture.  In the multimode case one cannot count on such a correspondence.  This is already apparent from the fact that, in the Heisenberg picture, the magnitude of the phase depends on the local pulse intensity (as in Eq.~(\ref{heissol2})), whereas the Schr\" odinger picture treatment determines a single value for the phase shift, simply by projecting the final field state onto the initial one, as in Eq.~(\ref{proj}).

It seems legitimate to say that spontaneous emission is ultimately responsible by the impossibility to get large phase shifts; in the ``slow regime,'' as just described, by removing a photon from the system, and in the ``fast regime'' by populating all the initially empty pairs of temporal modes (with the same $\mu$) with equal probability.  This is consistent with many previous results that indicate that in order to carry a nontrivial quantum logical operation (i.e., one that can change a state into an orthogonal one) with an error probability $P_e$ one needs of the order of $1/P_e$ photons, since $1/\bar n$ is precisely the ratio of ``spontaneous emission noise'' to ``signal,'' when one has $\bar n$ control photons \cite{jgb}.  In other words, single-photon quantum optical gates are bound to have failure probabilities of the order of unity.  This is plainly the case in schemes such as ``linear optics quantum computing'' \cite{kok}, which, however, have the advantage, over the schemes considered here, of not modifying the shape of the pulses when they succeed.

In view of all the evidence gathered thus far, and the possible pitfalls of incomplete analyses, it seems reasonable to suggest that any future proposals of ``single-photon Kerr nonlinearities'' for quantum logic should at least include detailed studies involving: (1) clearly local, and physically realizable Hamiltonians; (2) localized wavepackets, described by quantized multimode fields, where at least enough modes are included in numerical calculations to cover the whole nonlinear medium's bandwidth; (3) conventional fidelities computed in the Schr\" odinger picture; and (4) a realistic estimate of any residual losses or decoherence mechanisms.  As the results presented here indicate, however, there seems to be no reason to believe that such an analysis could violate the conclusions of Shapiro's analysis \cite{shapiro}.

This research has been supported by the National Science Foundation.

\end{document}